\documentclass{SCGE}

\begin{document}

\begin{picture}(0,0){\rm
\put(0,-20){\makebox[160truemm][l]{\bf {\sanhao\raisebox{2pt}{.}}
Article  {\sanhao\raisebox{1.5pt}{.}}}}}
\put(0,-34){\jiuwuhao {\textcolor[rgb]{0.5,0.5,0.5}{\sf
}}}
\end{picture}

%%%%%%%%%%%%%%%%%%%%%%%%%%%%%%%%%%%%%%%%%%%%%%%%%%%%%%%%%%%%
%\input psfig.sty
\def\bm{\boldsymbol}
\def\dl{\displaystyle}
\def\du{\end{document}}
\def\d{{\rm d}}
\def\e{{\rm e}}
\def\i{{\rm i}}
\renewcommand{\baselinestretch}{1.08} \baselineskip 12.2pt\parindent=10.8pt
\renewcommand{\thefootnote}
%%%%%%%%%%%%%%%%%%%%%%%%%%%%%%%%%%%%%%%%%%%%%%%%%%%%%%%%%%%%

% The author doesn't need fill in it.
\Year{2015} %
\Month{July} %
\Vol{58} %
\No{7} %
\BeginPage{070302} %
%\EndPage{} %
\AuthorMark{{\rm SEO}, et al.}  %
\AuthorMarkCite{{\rm SEO K, TIAN L}, et al.} %
\DOI{10.1007/s11433-015-5660-0} 
% The author doesn't need fill in it.

%%%%%%%%%%%%%%%%%%%%%%%%%%%%%%%%%%%%%%%%%%%%%%%%%%%%%%%%%%%%
% \title[short text for running head]{full title}{comments for title}
\title{Mott insulator-superfluid phase transition in a detuned multi-connected Jaynes-Cummings lattice}

\author{SEO Kangjun}{}
\author{TIAN Lin}{Corresponding author (email: ltian@ucmerced.edu)}
%\footnote{*Corresponding author (email: ldlzhangyu@yahoo.com.cn)}
\address[]{School of Natural Sciences, University of California, Merced, California 95343, USA;}

\maketitle \vspace{-3.5mm}{\footnotesize\begin{center} Received Month date, Year; accepted Month date, Year
\end{center}}\vspace*{-5mm}

% Abstract is required.
\begin{center}
\rule{16.5cm}{0.4pt}
\parbox{16.5cm}
{\begin{abstract}
The connectivity and tunability of superconducting quantum devices provide a rich platform to build quantum simulators and study novel many-body physics. Here we study quantum phase transition in a detuned multi-connected Jaynes-Cummings lattice, which can be constructed with superconducting circuits. This model is composed of alternatively connected qubits and cavity modes. Using a numerical method, we show that by varying the detuning between the qubits and the cavities, a phase transition from the superfluid phase to the Mott insulator phase occurs at commensurate fillings in a one-dimensional array. We study the phase transition in lattices with symmetric and asymmetric couplings, respectively. 
\end{abstract}}
\end{center}\vspace*{-0.6cm}

\begin{center}
\parbox{16.5cm}
{\bf\jiuhao Superconducting quantum devices, Quantum simulation, Jaynes-Cummings lattice, Phase transition}
\end{center}

\begin{center}
{}
%\CITA
\Cit{Seo K, Tian L. Mott insulator-superfluid phase transition in a detuned multi-connected Jaynes-Cummings lattice. Sci China-Phys Mech Astron, 2015, 57: 1--6, doi: }
\end{center}
%%%%%%%%%%%%%%%%%%%%%%%%%%%%%%%%%%%%%%%%%%%%%%%%%%%%%%%%%%%%

%%%%%%%%%%%%%%%%%%%%%%%%%%%%%%%%%%%%%%%%%%%%%%%%%%%%%%%%%%%%
\textwidth=178truemm 
\textheight=236truemm
\voffset=15mm
\wuhao\vspace*{8mm}
\begin{multicols}{2}
%%%%%%%%%%%%%%%%%%%%%%%%%%%%%%%%%%%%%%%%%%%%%%%%%%%%%%%%%%%%

\section{Introduction\label{sec:intro}}
Quantum simulation has become a frontier in quantum information science after it was first studied in the pioneer work of Feynman~\cite{feynman:82}. State-of-the-art technologies in building atomic and solid-state devices in the quantum limit enable us to build quantum simulators with rich controllability and novel physics effects~\cite{AshhabRMP2014}. Among such efforts, superconducting quantum simulation is being intensively explored in recent theoretical and experimental works. With superconducting circuits that contain only one or two qubits, people have demonstrated Anderson localization, switch of the Chern number in a topological phase transition, and the simultaneous coupling of a superconducting resonator to multiple qubits~\cite{suexp1,suexp2,suexp3,suexp4}. These experiments are enabled by technological developments in controlling and coupling superconducting devices with long decoherence times and high gate fidelity in the past decade~\cite{squbitReview, Barends:2013, Chow:2014,Reed:2012, Chen:2014, Srinivasan:2011}. Meanwhile, the experimental progress is paralleled by theoretical proposals on building both digital and analog quantum simulators with superconducting circuits. The many-body models being studied include quantum spin systems~\cite{Garcia-RipollPRB2008, Tian:2010, MarquardtPRL2013, GellerPRA2013, Zhang:2014, SolanoPRL2014}, models bearing novel topological properties~\cite{NoriPRA2010, JQY2011, GreentreePRL2012, Schmidt:2013}, models of electron-phonon interaction~\cite{Mei:2013,Stojanovic:2014}, and systems in high energy physics~\cite{Kapit:2013, Marcos:2013, PeropadrePRB2013}. One particularly interesting model is the so-called coupled cavity array (CCA) model that is made of arrays of optical or microwave cavities, in which each cavity couples to a nonlinear medium. It was shown that the CCA model can demonstrate Mott insulator (MI)-to-superfluid (SF) phase transition for cavity polaritons~\cite{Hartmann:2006,Greentree:2006,Angelakis:2007,Koch:2009,Na:2008,Hartmann:2008}, in analogy to the Bose-Hubbard model~\cite{Fisher:1989,Batrouni:1990}. One advantage of analog quantum simulators over general-purposed quantum computers is that they put less stringent requirements on the quantum logic operations and are more robust against decoherence and leakage errors. 

In a recent work, we proposed a multi-connected Jaynes-Cummings (JC) lattice model, which consists of arrays of alternatively coupled qubits and cavity modes~\cite{seo:2014-1}. This model can be constructed with superconducting systems, e.g., by connecting X-mon qubits and coplanar waveguide resonators~\cite{Barends:2013}. Using the exact diagonalization method, we studied the quantum phase transition of this model in a finite-size one-dimensional (1D) lattice. This system demonstrates phase transition from the incompressible MI phase to the gapless SF phase by varying the ratio between the left and the right qubit-cavity couplings. Moreover, the system shows the novel feature of reentrance to the MI phase from the SF phase when the coupling ratio is further increased. This reentrant behavior originates from the left-right symmetry between the couplings, and it distinguishes the multi-connected lattice model from the CCA. Note that effective cavity mode coupling and quantum magnetism were studied in interconnected qubit-cavity arrays that bear uniform or opposite couplings~\cite{QiuPRA2014, GarciaRipollPRL2014, GarciaRipollarXiv2014}.

Here we study a detuned multi-connected JC lattice, exploring the role of qubit-cavity detuning in the MI-to-SF phase transition. We show that the effective onsite nonlinearity strongly depends on the detuning. By varying the detuning from large negative to large positive values, the nonlinearity can be tuned from weak to strong, compared with the magnitudes of the qubit-cavity couplings. We use the exact diagonalization method to find the many-body ground state of a 1D lattice. Then, the single-particle density matrix at commensurate fillings is calculated. Note that even though 1D bosonic systems do not possess long range order, the single-particle density matrix decays much faster in the MI phase than in the SF phase and still gives clear signature of the phase transition. We study the quantum phases for lattices with symmetric and asymmetric couplings, respectively. For symmetric couplings with the left and right couplings equal to each other, the phase transition demonstrates universal behavior that only depends on the ratio between the detuning and the coupling; whereas for asymmetric couplings, the phase transition shows more complicated behavior. 

The qubit-cavity detuning can be conveniently adjusted by applying a global dc magnetic field to the SQUID loop of the X-mon qubits~\cite{Barends:2013}. The magnetic field controls the energy level splitting of the qubits, and hence the detuning, over a very wide range. In comparison to varying the qubit-cavity couplings, a practical advantage of varying the detuning is to avoid additional circuit elements that could bring in serious complications in circuit design and operation. 

This paper is organized as follows. In Sec.~\ref{sec:multiJC}, we present the Hamiltonian for a multi-connected JC lattice model in a 1D array, and we analyze the effective onsite interaction and the effective hopping matrix element. The dependence of the onsite interaction on the qubit-cavity detuning is studied in detail. Sec.~\ref{sec:detuning} starts with a description of our numerical method. We then present our results of the single-particle density matrix for both symmetric and asymmetric qubit-cavity couplings. Conclusions are given in Sec.~\ref{sec:concl}.

\section{Multi-connected JC lattice\label{sec:multiJC}}
\subsection{Model\label{ssec:model}}
\begin{figure}[H]
\centering
\includegraphics[width=\linewidth, clip]{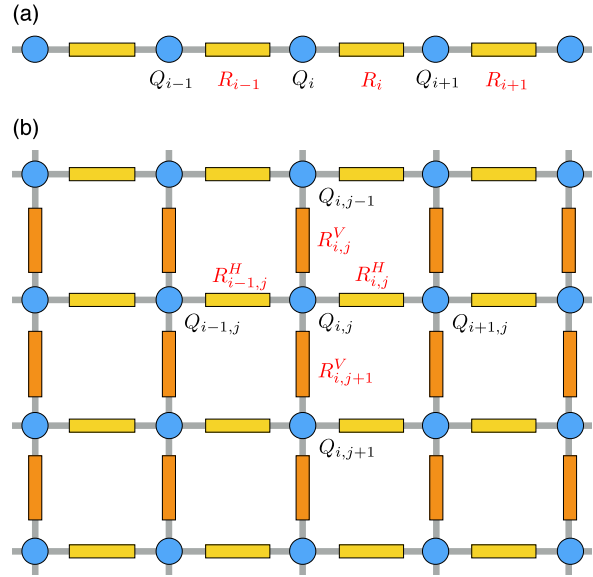}
\caption{Multi-connected JC lattice models. (a) A 1D lattice made of alternatively connected qubits $Q_{i}$ and cavities $R_{i}$. Each qubit $Q_i$ couples simultaneously to $R_{i}$ with coupling $g_r$ and to $R_{i-1}$ with coupling $g_l$. (b) A two-dimensional lattice with the qubits and the cavities connected in a checkerboard pattern. Here a unit cell consists of a qubit $Q_{i,j}$, a cavity $R_{i,j}^H$, and another cavity $R_{i,j}^V$. The qubit $Q_{i,j}$ couples to four cavities: $R_{i,j}^H$, $R_{i,j}^V$, $R_{i-1,j}^H$, and $R_{i,j+1}^V$, with respective couplings.}
\label{fig1}
\end{figure}

The multi-connected JC lattice model is made of alternatively connected qubits and cavities~\cite{seo:2014-1}, as illustrated in Fig.~\ref{fig1}(a). In a 1D configuration, each qubit couples to two neighboring cavities, one to the left and the other to the right hand side, with independent coupling constants $g_l$ and $g_r$. We define a unit cell as one qubit and the cavity to its right hand side. The local Hamiltonian for the $i$-th unit cell has the form of a standard JC model with ($\hbar = 1$)
\begin{equation}
H_{JC}^{i} = \omega_c a_i^\dag a_i + \frac{\omega_z}{2} \sigma_i^z + g_r \left( a_i^\dag \sigma_i^- + \sigma_i^+ a_i \right),\label{eq:intraham}
\end{equation}
where $\sigma_i^{\pm,z}$ denote the Pauli operators of the qubit, $a_i^\dag$ ($a_i$) is the creation (annihilation) operator of the cavity mode, $\omega_c$ is the angular frequency of the cavity, and $\omega_z$ the energy level splitting of the qubit. The qubit-cavity detuning $\Delta$ is defined as the frequency difference between the qubit and the cavity with $\Delta=\omega_c - \omega_z$.

In addition, the qubit at site $i$ couples to the cavity at site $i-1$ with the interaction 
\begin{equation}
H_\text{int}^{i} =g_l \left( a_{i-1}^\dag \sigma_{i}^- + \sigma_i^+ a_{i-1} \right),\label{eq:interham}
\end{equation}  
which is also in the form of the JC coupling. Combining the local Hamiltonian (\ref{eq:intraham}) and the interaction (\ref{eq:interham}), we can write the total Hamiltonian of the multi-connected JC lattice model as 
\begin{equation}
H_t = \sum_{i} \left( H_{JC}^i + H_\text{int}^{i} \right),\label{eq:Ht}
\end{equation}
where the summation runs over the entire lattice. It is noteworthy that the total Hamiltonian is invariant under a left-right reflection of the lattice accompanied by an exchange of the couplings $g_l$ and $g_r$. This reflection symmetry is at the heart of the reentrant behavior studied in our previous work~\cite{seo:2014-1}. Multi-connected lattices with more complicated geometries in higher dimension can also be studied. For example, the qubits and cavities can be connected in a checkerboard pattern in a two-dimensional lattice, as illustrated in Fig.~\ref{fig1}(b).

\subsection{Nonlinearity and hopping\label{ssec:Ut}}
The total Hamiltonian $H_t$ possesses the key elements for a MI-to-SF phase transition: an effective onsite nonlinear interaction and effective hopping matrix elements between neighboring sites. To illustrate this point, we study this model in the strongly-asymmetric limit of $g_l \ll g_r$ (or similarly, $g_{r}\ll g_{l}$). First, we consider the local Hamiltonian $H_{JC}^i$ at site $i$. The eigenstates of $H_{JC}^i$ include a zero-excitation state $|0_i\rangle = |0, \downarrow\rangle$ and the polariton doublets $|n_i, \pm\rangle$ with
\begin{subequations}
\begin{align}
|n_{i},+\rangle &= \cos \frac{\delta_{n_{i}}}{2} |n_{i},\downarrow\rangle+\sin \frac{\delta_{n_{i}}}{2} |n_{i}-1,\uparrow\rangle,\label{eq:n+} \\
|n_{i},-\rangle &= \sin \frac{\delta_{n_{i}}}{2} |n_{i},\downarrow\rangle-\cos \frac{\delta_{n_{i}}}{2} |n_{i}-1,\uparrow\rangle,\label{eq:n-}
\end{align}
\end{subequations}
where $\vert\uparrow, \downarrow\rangle$ correspond to the spin up or down states of the Pauli operator $\sigma_{i}^{z}$, the $n_{i}$ index on the right hand sides of the equations refers to cavity photon numbers, and the $n_i$ index in the eigenstates $|n_i, \pm\rangle$ refers to the total number of excitations at site $i$ that counts both spin flip and photon excitations. Here $\cos \frac{\delta_{n_{i}}}{2} = \sqrt{\frac{1}{2} (1 + \frac{\Delta}{\Omega_{n_{i}}} )}$, $\sin \frac{\delta_{n_{i}}}{2} = \sqrt{\frac{1}{2} (1 - \frac{\Delta}{\Omega_{n_{i}}})}$, and $\Omega_{n_{i}}=\sqrt{\Delta^2+4g_r^2n_i}$ is an effective Rabi frequency. The polariton doublets $|n_i,\pm_i\rangle$ are eigenstates with $n_i$ excitations. The eigenenergies can be written as 
\begin{equation}
\varepsilon_{n_i,\pm} = (n_i-1/2) \omega_c \pm \Omega_{n_i}/2\label{eq:een}
\end{equation}
and $\varepsilon_{0_i}=-\omega_z/2$, respectively. The states $|n_i, -\rangle$ have lower energy in the $n_{i}$ doublet, and are referred to as the lower polariton states. The energy difference $\Delta\varepsilon_{n_i} = \varepsilon_{n_i,-}-\varepsilon_{0_i}$ is thus the lowest energy cost to produce $n_i$ excitations at site $i$. 

\begin{figure}[H]
\centering
\includegraphics[width=1.0\linewidth, clip]{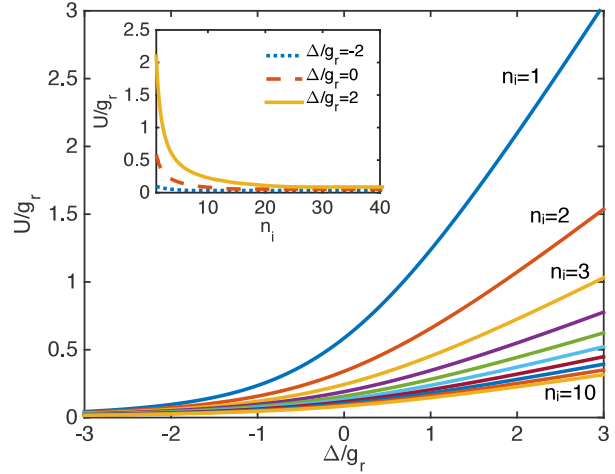}
\caption{The effective nonlinearity $U$ versus $\Delta/g_{r}$ with $g_r = 150$ MHz. Inset: $U$ versus $n_i$ for $\Delta/g_r = -2$, 0 and $2$, respectively.}
\label{fig2}
\end{figure}

To analyze the effective nonlinearity, we use the following estimation. Assume that the energy of lower polariton states $|n_{i},-\rangle$ can be described by an effective Hamiltonian $H_\text{eff} = \omega_p p_i^\dag p_i + (U/2) p_i^\dag p_i^\dag p_i p_i $, where $\omega_{p}$ is the single-polariton energy and $U$ is the magnitude of the Hubbard onsite interaction. The energy cost to generate $n_{i}$ excitations is thus $\Delta\varepsilon_{n_i} = n_i \omega_{p} + (U/2) n_i (n_i-1)$, with the special case of $\Delta \varepsilon_{1_i}=\omega_{p}$. With these relations, the Hubbard interaction can be written as $U=(\Delta\varepsilon_{n_{i+1}}-\Delta\varepsilon_{n_{i}}-\omega_{p})/n_{i}$. Combining this result with the eigenenergies given in (\ref{eq:een}), we can estimate the interaction $U$ in terms of the detuning and the qubit-cavity coupling for selected $n_{i}$ value. Note that different from the Hubbard model, the effective $U$ in our estimation depends on $n_{i}$. The JC model cannot be exactly mapped to the Hubbard model, even though the nonlinearity is a convenient way to analyze the behavior of this system. For example, at zero detuning with $\Delta=0$, $U=g_r(1 + \sqrt{n_i} - \sqrt{n_i+1})/n_i$ for $n_{i}$ excitations, proportional to the coupling strength $g_{r}$ and decreasing monotonically with the total excitation number $n_{i}$. The dependence of $U$ on $n_{i}$ is shown in the inset of Fig.~\ref{fig2}. 

In this system, the detuning can be an effective knob to tune the phase transition between the MI and the SF phases. Practically speaking, it is easier to control the qubit-cavity detuning than to adjust the qubit-cavity couplings in the laboratory. The detuning of superconducting qubits can be controlled with global external magnetic field over a very wide range~\cite{Barends:2013}. Below we study the dependence of the Hubbard $U$ on the detuning. In Fig.~\ref{fig2}, we plot $U$ as functions of the detuning for low-lying polariton states with $n_i=1$ to $10$ and a coupling strength of $g_r = 150$ MHz. The nonlinearity increases continuously with the detuning. In the limit of large detuning magnitude with $|\Delta| \gg g_0$, we find that $U\approx (\Delta + |\Delta|)/2n_i$, i.e., for large positive detuning, $U\rightarrow \Delta/n_{i}$ and for large negative detuning, $U\rightarrow 0$. This indicates that for low-lying polariton excitations, the SF phase is more favorable at negative detuning with a nearly vanishing effective interaction; and the MI phase is more favorable at large positive detuning. 

Assume the coupling $g_{l}$ between neighboring sites to be finite with $g_l \ll g_r$. The coupling $H_\text{int}^{i}$ between adjacent sites can be treated as a perturbation to the eigenstates (\ref{eq:n+}) and (\ref{eq:n-}). Consider the matrix element of $H_\text{int}^{i}$ between state $|A\rangle = |n_{i-1},- \rangle\otimes|n_i,- \rangle$, which contains $n_{i-1}$ excitations at site $i-1$ and $n_{i}$ excitations at site $i$, and state $|B\rangle = | n_{i-1}-1,-\rangle \otimes| n_i+1,-\rangle $, which includes $(n_{i-1}-1)$ excitations at site $i-1$ and $(n_{i}+1)$ excitations at site $i$. We find
\begin{eqnarray}
&&\langle B| \sigma_i^+ a_{i-1}|A \rangle = - \sin\frac{\delta_{n_i}}{2} \cos\frac{\delta_{n_{i}+1} }{2} \times  \nonumber \\
&& \left(\sqrt{n_{i-1}}\sin\frac{\delta_{n_{i-1}-1}}{2} \sin\frac{\delta_{n_{i-1}}}{2}\right. \nonumber \\
&&+ \left.\sqrt{n_{i-1}-1} \cos\frac{\delta_{n_{i-1}-1}}{2} \cos\frac{\delta_{n_i-1}}{2}\right),\label{eq:BmatrixA}
\end{eqnarray}
which can be viewed as a hopping matrix element for polariton excitations between neighboring sites, with a hopping amplitude proportional to the coupling $g_{l}$. The total number of excitations in the lattice is preserved during the hopping process. Hence in the limit of very asymmetric couplings with $g_l \ll g_r$, the onsite coupling $g_{r}$ produces an effective Hubbard $U$ and the weak nonlocal coupling $g_{l}$ produces a hopping term between adjacent lattice sites. Similar argument can be applied to the opposite limit of $g_{l}\gg g_{r}$, in which case $g_{r}$ becomes the hopping perturbation and $g_{l}$ yields the effective nonlinearity.

Compared with the CCA, the multi-connected JC lattice model bears the distinct feature that the cavities do not interact directly, and as a result, the hopping of the excitations is not caused by direct coupling between cavity modes. In the CCA, local qubit-cavity coupling produces onsite nonlinearity and cavity coupling induces hopping between neighboring sites. Whereas in the multi-connected JC lattice, both onsite nonlinearity and hopping are caused by the qubit-cavity couplings. In particular, when the magnitudes of the left and the right couplings are comparable ($g_{l}\sim g_{r}$), the role of each coupling cannot be simply analyzed as either inducing the nonlinearity or generating the hopping. It is the interplay between the left and the right couplings that give rise to the phase transition in this model.

\section{Detuning induced phase transition}\label{sec:detuning}
\subsection{Numerical method}
In this section, we study detuning-induced quantum phase transition in the multi-connected JC lattice model using the exact diagonalization method. For this purpose, we choose the basis vectors 
\begin{equation}
|\psi\rangle 
= |n_1,\sigma_1\rangle 
\otimes
|n_2,\sigma_2\rangle 
\cdots 
|n_M,\sigma_M\rangle
\end{equation}
to construct the matrix representation of the total Hamiltonian for a finite system with lattice size $M$ and total excitation number $N$. Here $n_{i}$ is photon excitations, and $\sigma_i=\downarrow,\uparrow$ is the qubit state at site $i$ with $\downarrow$ ($\uparrow$) corresponding to $0$ ($1$) spin excitation. The operator for the total excitations in the lattice is defined as $\hat{N}=\sum_i (a_i^\dag a_i +  \sigma_i^+ \sigma_i^-)$, including both photon and spin excitations. In our model, with the commutation relation $[\hat{N},\,H_{t}]=0$, the total excitation number is a good quantum number. Below we study the quantum phase of many-body ground state at commensurate fillings, i.e., the total excitation number $N$ is a multiple of the lattice size $M$. Hence, only these basis vectors that have $N$ excitations are selected in the numerical calculation. Because the Hamiltonian is a sparse matrix on the selected vector space, we use a Lanczos algorithm to solve the many-body ground state $|G\rangle$. Note that we have assumed the periodic boundary condition for the lattice to ensure translational symmetry.  

For a finite size system, the order parameter $\langle G | a_{i} | G \rangle$, which is often used to characterize quantum phase transition in the thermodynamic limit, is always equal to zero because of particle number conservation. Instead, we use the single-particle density matrix
defined as 
\begin{equation}
\rho_{1}(i,j) = \langle G| a_i^\dag a_j |G\rangle / \langle G| a_i^\dag a_i |G\rangle\label{eq:rho1}
\end{equation}
to characterize the phase transition in this model. From definition, $\rho_{1}(i,j)$ is hermitian and semi-positive definite. Furthermore, due to the translational and reflectional symmetry of the multi-connected JC lattice under periodic boundary condition, $\rho_{1}(i,j)$ is also real, symmetric, and cyclic. Hence $\rho_{1}(i,j)$ only depends on $|i-j|$, and hereafter we write the single-particle density matrix as $\rho_{1}(x)$ with $x=|i-j|$. For 1D bosonic systems in the SF phase, $\rho_{1}(x)$ decays algebraically; whereas in the MI phase, $\rho_{1}(x)$ decreases exponentially to zero. Hence, $\rho_{1}(x)$ in the SF phase could have much larger value at finite $x$ than that of the MI phase due to its slower decay. Note that due to the finite size effect, $\rho_{1}(x)$ in our calculation will not reach zero in the MI phase. It decreases to a small finite value that will be clearly distinguishable from that of the SF phase. As a result, we can verify the occurrence of the phase transition, even though we cannot accurately determine the position of the quantum critical point. 

In our calculation, we choose a lattice size $M=8$ and a total excitation number $N=8$, with the qubit-cavity couplings in a range of $0$ - $300$ MHz and the qubit-cavity detuning in the range of $\Delta/g_{0} \in [-3, 3]$. Here $g_{0}=150$ MHz is introduced as a unit of the detuning, with its strength comparable to the qubit-cavity couplings. 

\subsection{Phase transition with symmetric couplings\label{ssec:symcoupling}}
\begin{figure}[H]
\centering
\includegraphics[width=1.0\linewidth, clip]{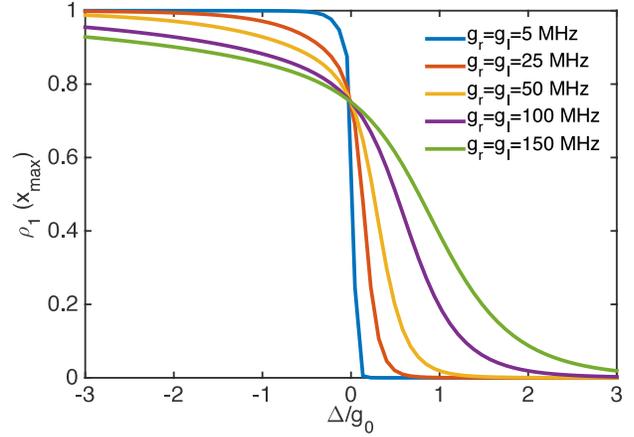}
\caption{The single-particle density matrix $\rho_1(x=x_\text{max})$ versus the unitless detuning $\Delta / g_0 $ at selected couplings with $g_l=g_r$ and $g_0 = 150$ MHz.}
\label{fig3}
\end{figure}

First, we study the detuning effect on the quantum many-body phase of a symmetric multi-connected JC lattice model, in which the left and the right qubit-cavity couplings are equal to each other with $g_l=g_r$. The many-body phase of this system thus only depends on the ratio between the detuning and the coupling $g_{r}$. In Fig.~\ref{fig3}, we plot the single particle density matrix $\rho_{1}(x=x_\text{max})$ as a function of the detuning $\Delta$ at the maximal lattice distance $x_\text{max}$ and at selected values of the coupling $g_{r}$. Note that for the lattice size $M=8$, $x_\text{max}=4$ under the periodic boundary condition.

The numerical result shows that $\rho_1(x_\text{max})$ decreases monotonically with the detuning. For large negative detuning with $-\Delta \gg g_{r}$,  $\rho_1(x_\text{max})\to 1$. At $\Delta=0$, $\rho_1(x_\text{max})\sim 0.75$, which is still a large finite value of the order of unity. For large positive detuning with $\Delta \gg g_{r}$, however, $\rho_1(x_\text{max})\to 0$. The system hence exhibits a transition from the SF to the MI phases as $\Delta$ increases from negative to positive values. This result shows excellent agreement with the nonlinearity given in Fig.~\ref{fig2}, in which the effective nonlinearity increases from nearly zero to large finite value as the detuning increases. The phase transition can be explained with an intuitive physical picture. For large negative detuning, $\omega_{c} \ll \omega_{z}$, and the lower polariton states can be approximated as photon states with equal energy spacing, i.e., having weak nonlinearity. The many-body ground state in the presence of finite qubit-cavity coupling is then a superfluid of the photons. For large positive detuning, on the other hand, $\omega_{c} \gg \omega_{z}$, and the lower polariton states are approximately the qubit states that are localized at its own site. The many-body ground state then resembles the MI phase where hopping is prohibited by large nonlinearity.

Note that for symmetric couplings, the behavior of this system depends on the ratio $\Delta/g_{r}$ only. Hence the curves for different $g_{r}$ values in Fig.~\ref{fig3} can be rescaled to become one single curve. In other words, the width of the transition region between the SF and the MI phases scales with the coupling constant $g_{r}$, with wider detuning region for greater $g_{r}$ value. It is also worth noting that $\rho_1(x_\text{max})$ at $\Delta = 0$ is independent of the coupling $g_r$ with the system in the SF phase regardless of the coupling strength.

\subsection{Phase transition with asymmetric couplings\label{ssec:asymcoupling}}
\begin{figure}[H]
\centering
\includegraphics[width=1.0\linewidth, clip]{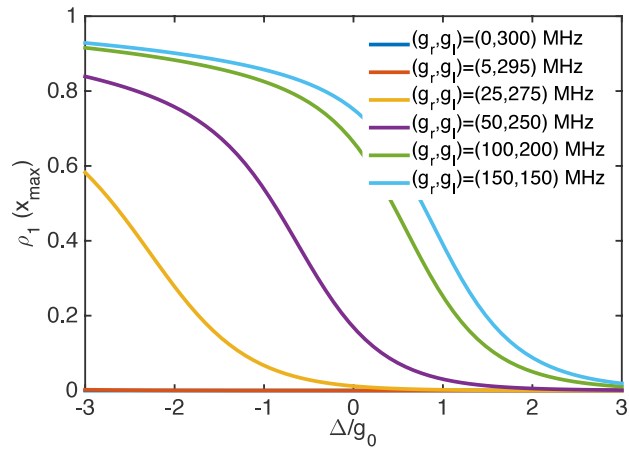}
\caption{The single-particle density matrix $\rho_1(x=x_\text{max})$ versus the unitless detuning $\Delta / g_0 $ at selected pairs of couplings and $g_0 = 150$ MHz. The couplings are $(g_l,g_r) = (0,300),\,(5,295),\,(25, 275),\,(50,250),\,(100,200)$, and $(150,150)$ MHz, respectively.}
\label{fig4}
\end{figure}

Now we consider the detuning effect on the phase transition of asymmetric systems with distinct qubit-cavity couplings $g_l$ and $g_r$. We calculate $\rho_1(x_\text{max})$ as a function of $\Delta$ for selected pairs of couplings as shown in Fig.~\ref{fig4}. The couplings are chosen in the experimentally approachable regime of $g_{l,r}\in[0,\, 300]$ MHz and satisfy $g_l+g_r = 300$ MHz. The numerical result shows the generic feature that the single-particle density matrix $\rho_1(x_\text{max})$ decreases with the increase of the detuning. In contrast to the case of symmetric couplings in Sec.~\ref{ssec:symcoupling}, the ratio of the qubit-cavity coupling $g_l/g_r$ plays an important role in the many-body phase of the ground state. For instance, at $(g_l,g_r) = (5,295)$ ($g_{l}/g_{r}=0.017$), $\rho_1(x_\text{max})\approx 0$ in the entire region of $\Delta/g_0\in [-3,\,3]$, which indicates that the system is in the MI phase. Whereas at $(g_l,g_r) = (50,250)$ ($g_{l}/g_{r}=0.2$), $\rho_1(x_\text{max})>0.8$ for $\Delta/g_0 = -3$, clearly indicating that the system is in the SF phase. Hence depending on the ratio between the couplings, the phase transition can occur at either negative or positive detuning. 

\section{Conclusions\label{sec:concl}}
To conclude, we studied the quantum phase transition in the many-body ground state of a 1D multi-connected JC lattice model. Using a numerical method, we showed that by varying the qubit-cavity detuning, the SF-to-MI phase transition can be observed in this system. The quantum critical point not only depends on the magnitude of the couplings, but also depends on the ratio between the left and the right couplings of the lattice. Our results show that the multi-connected JC lattice model could be a rich system to explore the many-body physics of cavity polaritons. 

\vspace*{2mm} 
\Acknowledgements{\bahao This work is supported by the National Science Foundation under Award Number 0956064.}

\end{multicols}

\end{document}